\documentclass[%
 reprint,
superscriptaddress,
nofootinbib,
 amsmath,amssymb,
 aps,
floatfix,
]{revtex4-2}
\usepackage{tikz}
\usepackage{braket}
\usepackage{bbm}
\usepackage{graphicx}
\usepackage{dcolumn}
\usepackage{bm}
\usepackage{hyperref}

\newcommand{\beginsupplement}{%
        \setcounter{table}{0}
        \renewcommand{\thetable}{S\arabic{table}}%
        \setcounter{figure}{0}
        \renewcommand{\thefigure}{S\arabic{figure}}%
     }
\newcommand{\Tr}{\text{Tr}}

\begin{document}

\preprint{APS/123-QED}

\title{Relative Entropy of Random States and Black Holes}

\author{Jonah Kudler-Flam}
\email{jkudlerflam@uchicago.edu}
 \affiliation{Kadanoff Center for Theoretical Physics, University of Chicago, Chicago, IL 60637, USA}



\date{\today}

\begin{abstract}


We study the relative entropy of highly excited quantum states. First, we sample states from the Wishart ensemble and develop a large-N diagrammatic technique for the relative entropy. The solution is exactly expressed in terms of elementary functions. We compare the analytic results to small-N numerics, finding precise agreement. Furthermore, the random matrix theory results accurately match the behavior of chaotic many-body eigenstates, a manifestation of eigenstate thermalization. We apply this formalism to the AdS/CFT correspondence where the relative entropy measures the distinguishability between different black hole microstates. We find that black hole microstates are distinguishable even when the observer has arbitrarily small access to the quantum state, though the distinguishability is nonperturbatively small in Newton's constant. Finally, we interpret these results in the context of the subsystem Eigenstate Thermalization Hypothesis (sETH), concluding that holographic systems obey sETH up to subsystems half the size of the total system.


\end{abstract}

\maketitle



\paragraph*{Introduction.}

Random matrices are a unifying subject in quantum physics. From encoding quantum information \cite{2016JMP....57a5215C}, to characterizing complicated many-body systems and quantum chaos \cite{2016AdPhy..65..239D}, to serving as toy models of the black hole information problem \cite{1993PhRvL..71.3743P,2007JHEP...09..120H}, random quantum states have become invaluable across many distinct subfields. Moreover, the mathematical field of random matrix theory is very mature, enabling analytical calculations in random states that would be otherwise intractable.

With the broad motivations of understanding structural properties of density matrices, quantum thermalization in isolated many-body systems, and the black hole information problem, we study the relative entropy of random quantum states. The applicability of this study to elucidating our motivating principles will subsequently be made clear.

The relative entropy between two density matrices $\rho$ and $\sigma$ is defined as
\begin{align}
    D(\rho \lvert \rvert \sigma ) := \Tr \left[ \rho\left( \log \rho - \log \sigma\right) \right].
\end{align}
As a distinguishability measure, it obeys various nice properties, such as positivity with $ D(\rho \lvert \rvert \sigma ) = 0$ if and only if $\rho = \sigma$. Crucially, the relative entropy is monotonic under quantum operations \cite{lindblad1975}
\begin{align}
    D(\mathcal{N}(\rho) \lvert \rvert \mathcal{N}(\sigma) ) \leq D(\rho \lvert \rvert \sigma ),
\end{align}
where $\mathcal{N}$ is any completely-positive trace-preserving map. A particularly important quantum operation that we will come back to is the partial trace. Monotonicity in this context means that density matrices become less distinguishable as you throw out more information about them, an intuitive notion.

Relative entropy is truly the mother of all quantities in quantum information theory. While at face value, it just measures the distinguishability between two density matrices, upon further inspection, its fundamental properties underlie many of the deepest universal statements about quantum mechanics \cite{lindblad1974,2002RvMP...74..197V}, quantum field theory \cite{2004PhLB..600..142C,2016JHEP...09..038F,2019JHEP...09..020B}, and quantum gravity \cite{2008CQGra..25t5021C,2012PhRvD..85j4049W}. 

While this progress has been significant, we will show that relative entropy has quite a bit more to tell us about each of these subdisciplines.
\begin{enumerate}
    \item[$\star$] By finding a closed form solution for the relative entropy of random density matrices, we characterize the space of quantum states. In the language of quantum hypothesis testing \cite{cmp/1104248844, 2005atqs.book...28O}, this precisely determines the error that one achieves for a measure one set of quantum states when performing a hypothesis test with limited access to the quantum state.
    \item[$\star$] While our general formula is exact in the limit of large Hilbert space dimensions, we find it to be remarkably accurate even for small Hilbert space dimensions. More interestingly, we find it to accurately predict the behavior of relative entropy between eigenstates of chaotic many-body Hamiltonians. These numerical observations imply that our results may be observable in noisy intermediate-scale quantum (NISQ) technologies \cite{2018arXiv180100862P}.
    \item[$\star$] Through applying our formalism to holographic quantum field theories, we conclude that the relative entropy between subregions of black hole microstates is finite, though non-perturbatively small in Newton's constant ($G_N$) up until the subregion is half of the total system size. Using quantum information inequalities, we show that this implies an extremely strong version of the eigenstate thermalization hypothesis  \cite{1991PhRvA..43.2046D,1994PhRvE..50..888S}. 
\end{enumerate}


\paragraph*{Random Mixed States.}

We begin with a Haar random pure state on a bipartite Hilbert space $\mathcal{H}_A\otimes \mathcal{H}_B$ \cite{2009arXiv0910.1768C}
\begin{align}
    \ket{\Psi} = \sum_{i=1}^{d_A} \sum_{\alpha=1}^{d_B} X_{i\alpha}\ket{i}_A \otimes \ket{\alpha}_B,
\end{align}
where the states in the sum are orthonormal bases for the sub-Hilbert spaces of dimensions $d_A$ and $d_B$ which we will always assume to be independently large.  The $X_{i\alpha}$'s are independently distributed complex Gaussian random variables with joint probability distribution \cite{2009arXiv0910.1768C,2001JPhA...34.7111Z}
\begin{align}
    P(\{ X_{i\alpha}\}) = \mathcal{Z}^{-1}\exp\left[-d_Ad_B \Tr \left(X X^{\dagger}\right)\right],
\end{align}
where $\mathcal{Z}$ is the normalization constant, ensuring the expression defines a probability. Here, $X$ represents the rectangular matrix whose matrix elements in the $i,\alpha$ basis are $X_{i\alpha}$. The random induced states on $\mathcal{H}_A$ are then
\begin{align}
    \rho_A = \frac{XX^{\dagger}}{\Tr(X X^{\dagger})}.
\end{align}
We note that the denominator is a random variable that is sharply peaked around unity, so we can ignore it in the limit of large Hilbert space dimension \cite{2009arXiv0910.1768C,2001JPhA...34.7111Z}
\begin{align}
    \rho_A \simeq XX^{\dagger}.
\end{align}
This defines the Wishart ensemble. We now introduce a diagrammatic representation of the density matrix \cite{1995NuPhB.453..531B,2008AcPPB..39..799J,2020arXiv201101277S}
\begin{align}
    \left[\ket{\Psi}\bra{\Psi}\right]_{i\alpha,j\beta}= X_{i\alpha} X^\ast_{j\beta}
    =
    \,
    \tikz[baseline=-0.5ex]{
    \draw[dashed] (0,0.2) node[align=center, above] {\footnotesize $\alpha$} -- (0,-0.2);
    \draw[dashed] (1,0.2) node[align=center, above] {\footnotesize $\beta$} -- (1,-0.2);
    \draw (-0.2,0.2) node[align=center, above] {\footnotesize $i$} -- (-0.2,-0.15);
    \draw (1.2,0.2) node[align=center, above] {\footnotesize $j$} -- (1.2,-0.15);
    }\ .
\end{align}
The solid and dashed lines correspond to subsystems $A$ and $B$ respectively. Matrix manipulations are done at the bottom edge of the diagram. For example, the partial trace over $\mathcal{H}_B$ is 
\begin{align}
    \label{eq:rho_diag}
    [\rho_A]_{i,j}= 
    {\sum_{\alpha=1}^{d_B} X_{i\alpha} X^\ast_{j\alpha}}
:= 
\,
    \tikz[baseline=-0.5ex]{
    \draw[dashed] (0,0.2) node[align=center, above] {\footnotesize $\alpha$} -- (0,0);
    \draw[dashed] (0,0)  -- (1,0);
    \draw[dashed]  (1,0.2) node[align=center, above] {\footnotesize $\alpha$} -- (1,0);
    \draw (-0.2,0.2) node[align=center, above] {\footnotesize $i$} -- (-0.2,-0.15);
    \draw (1.2,0.2) node[align=center, above] {\footnotesize $j$} -- (1.2,-0.15);
    }\ .
\end{align}
Ensemble averaging is done at the top of the diagram with propagators carrying  weight
\begin{align}
    \,
    \tikz[baseline=0ex]{
    \draw[dashed] (1.0,0) arc (0:180:0.5);
    \draw (1.2,0.) arc (0:180:0.7);
    }\
    := \braket{X_{i\alpha} X^\ast_{j \beta}}=  \frac{1}{d_A d_B}\, \delta_{ij} \delta_{\alpha\beta}.
    \label{eq:doubleline1}
\end{align}
Putting these operations together, we can, for example, take the trace of the density matrix
\begin{align}
    \braket{\Tr\rho_A}=
    \,
    \tikz[baseline=0ex]{
    \draw[dashed] (1.0,0) arc (0:180:0.5);
    \draw[dashed] (0,0) -- (1,0);
    \draw (1.2,0.) arc (0:180:0.7);
    \draw (-0.2,0.0)-- (-0.2,-0.15)--(1.2,-0.15)-- (1.2,0.0);
    }\ = 1,
\end{align}
where every closed loop gives a factor of the Hilbert space dimension. The diagrammatic rules for averaging assert that we must sum over all possible contractions of the bras and kets. For relative entropy, we need two independent density matrices, $\rho_A$ and $\sigma_A$. These must be averaged over the ensemble separately. To make this distinction, we color $\sigma_A$ red.

The logarithms in the definition of relative entropy make the quantity significantly more difficult to compute analytically than simple powers of the density matrices. Happily, a replica trick for the relative entropy has been developed that re-expresses the logarithm as a limit of appropriate powers \cite{2016PhRvL.117d1601L}
\begin{align}
    D(\rho \lvert \rvert \sigma) = \lim_{n\rightarrow1}\frac{1}{n-1} \left(\log \Tr \rho^n- \log \Tr \rho  \sigma^{n-1}\right).
    \label{S_replica}
\end{align}
We will compute these two terms separately. The first term is recognized as minus the R\'enyi entropy. For $n = 2$, we have
\begin{align}
\label{eq:tr_r2}
    \Tr\rho_A^2= 
    \,
    \tikz[baseline=0ex]{
    \draw[dashed] (0,0.2)-- (0,0);
    \draw[dashed] (0,0) -- (1,0);
    \draw[dashed]  (1,0.2) -- (1,0);
    \draw (-0.2,0.2)-- (-0.2,-0.25);
    \draw (1.2,-0.1)-- (1.2,0.2);
    \draw[dashed] (2,0.2)-- (2,0);
    \draw[dashed] (2,0) -- (3,0);
    \draw[dashed]  (3,0.2) -- (3,0);
    \draw (1.8,0.2)-- (1.8,-0.1);
    \draw (3.2,-0.25)-- (3.2,0.2);
    \draw (1.2,-0.1)-- (1.8,-0.1);
    \draw (-0.2,-0.25) -- (3.2,-0.25);
    }\ .
\end{align}
The ensemble average is a sum of the two contractions
\begin{align}
\label{purity_diagram}
    \braket{\Tr\rho_A^2} &=
    \,
    \tikz[scale=0.8,baseline=0.5ex]{
    \draw[dashed] (0,0) -- (1,0);
    \draw (-0.2,0.)-- (-0.2,-0.25);
    \draw (1.2,-0.1)-- (1.2,0.);
    \draw[dashed] (2,0) -- (3,0);
    \draw (1.8,0.)-- (1.8,-0.1);
    \draw (3.2,-0.25)-- (3.2,0.);
    \draw (1.2,-0.1)-- (1.8,-0.1);
    \draw (-0.2,-0.25) -- (3.2,-0.25);
    \draw[dashed] (1.0,0) arc (0:180:0.5);
    \draw[dashed] (3.0,0) arc (0:180:0.5);
    \draw (1.2,0) arc (0:180:0.7);
    \draw (3.2,0) arc (0:180:0.7);
    }
\ \,
+
\ \,
    \tikz[scale=0.7,baseline=0.5ex]{
    \draw[dashed] (0,0) -- (1,0);
    \draw (-0.2,0.)-- (-0.2,-0.25);
    \draw (1.2,-0.1)-- (1.2,0.);
    \draw[dashed] (2,0) -- (3,0);
    \draw (1.8,0.)-- (1.8,-0.1);
    \draw (3.2,-0.25)-- (3.2,0.);
    \draw (1.2,-0.1)-- (1.8,-0.1);
    \draw (-0.2,-0.25) -- (3.2,-0.25);
    \draw[dashed] (2.0,0) arc (0:180:0.5);
    \draw[dashed] (3.0,0) arc (0:180:1.5);
    \draw (1.8,0) arc (0:180:0.3);
    \draw (3.2,0) arc (0:180:1.7);
    },
\ \,
\end{align}
immediately giving $d_A^{-1} + d_B^{-1}$. This can be generalized to arbitrary powers. Because of the sum over all possible contractions, in general, the moments are expressible as a sum over the permutation group
\begin{align}
    \langle \Tr \rho_A^n \rangle = \frac{1}{(d_A d_B)^n} \sum_{\tau \in S_n} d_A^{D(\eta^{-1} \circ \tau)}d_B^{D(\tau)},
    \label{renyi_sum}
\end{align}
where $D(\cdot)$ is the number of cycles in the permutation and $\eta$ is the cyclic permutation. Each permutation corresponds to a diagram with the cycle structure determining which bra is contracted with which ket. For example, in \eqref{purity_diagram}, the first diagram corresponds to the identity permutation because each bra is contracted with its own ket while the second diagram corresponds to the swap permutation because the bra of the first density matrix is contracted with the ket of the second and vice versa. These are the only elements of $S_2$.

When the Hilbert spaces are large, only the terms that maximize $D(\eta^{-1} \circ \tau) + D( \tau)$ will contribute to the sum at leading order. These are known as the non-crossing permutations and have $D(\eta^{-1} \circ \tau) + D( \tau) = n+1$. Much is known about this special subset of permutations including that the number of such permutations with $D(\eta^{-1} \circ \tau) = k$ is given by the Narayana number \cite{KREWERAS1972333, SIMION2000367}
\begin{align}
    N_{n,k} = \frac{1}{n}\binom{n}{k}\binom{n}{k-1}.
\end{align}
Thus, the sum can be reorganized as
\begin{align}
    \langle \Tr \rho_A^n \rangle &= \frac{1}{(d_A d_B)^n} \sum_{k = 1}^n N_{n,k} d_A^{k}d_B^{n+1-k}
    \nonumber
    \\
    &= d_A^{1-n} \, _2F_1\left(1-n,-n;2;\frac{d_A}{d_B}\right),
    \label{renyi_hyper}
\end{align}
where $\, _2F_1$ is a hypergeometric function. This reproduces Page's famous result \cite{1993PhRvL..71.1291P}.

Next, we consider the second term of \eqref{S_replica}. For simplicity, we first consider the overlap between the two density matrices which, as a diagram, looks like
\begin{align}
    \Tr (\rho_A \sigma_A) = 
    \,
    \tikz[baseline=0ex]{
    \draw[dashed] (0,0.2)-- (0,0);
    \draw[dashed] (0,0) -- (1,0);
    \draw[dashed]  (1,0.2) -- (1,0);
    \draw (-0.2,0.2)-- (-0.2,-0.25);
    \draw (1.2,-0.1)-- (1.2,0.2);
    \draw[dashed,red] (2,0.2)-- (2,0);
    \draw[dashed,red] (2,0) -- (3,0);
    \draw[dashed,red]  (3,0.2) -- (3,0);
    \draw[red] (1.8,0.2)-- (1.8,-0.1);
    \draw[red] (3.2,-0.25)-- (3.2,0.2);
    \draw (1.2,-0.1)-- (1.8,-0.1);
    \draw (-0.2,-0.25) -- (3.2,-0.25);
    }\ .
\end{align}
We must ensemble average the black and red lines separately, so there is only a single term
\begin{align}
    \braket{\Tr (\rho_A \sigma_A)} &=
    \,
    \tikz[scale=0.8,baseline=0.5ex]{
    \draw[dashed] (0,0) -- (1,0);
    \draw (-0.2,0.)-- (-0.2,-0.25);
    \draw (1.2,-0.1)-- (1.2,0.);
    \draw[dashed,red] (2,0) -- (3,0);
    \draw[red] (1.8,0.)-- (1.8,-0.1);
    \draw[red] (3.2,-0.25)-- (3.2,0.);
    \draw (1.2,-0.1)-- (1.8,-0.1);
    \draw (-0.2,-0.25) -- (3.2,-0.25);
    \draw[dashed] (1.0,0) arc (0:180:0.5);
    \draw[dashed,red] (3.0,0) arc (0:180:0.5);
    \draw (1.2,0) arc (0:180:0.7);
    \draw[red] (3.2,0) arc (0:180:0.7);
    },
\ \,
\label{overlap_diagram}
\end{align}
giving $d_A^{-1}$. We again generalize this to arbitrary powers by expressing the moments in terms of a sum over the permutation group
\begin{align}
    \langle \Tr (\rho_A\sigma_A^{n-1} )\rangle = \frac{1}{(d_A d_B)^n} \sum_{\tau \in \mathbbm{1}\times S_{n-1}} d_A^{D(\eta^{-1} \circ \tau)}d_B^{D(\tau)}.
    \label{relative_sum}
\end{align}
The crucial difference between this expression and \eqref{renyi_sum} is that the sum is only over a subgroup of permutations, namely the ones that stabilize the first element. The first element (black lines) must be stabilized because the black lines must be contracted with themselves and there is only a single element with black lines. The reason the swap permutation was not included in \eqref{overlap_diagram} is because it acts nontrivially on the first density matrix.

\begin{figure}
    \centering
    \includegraphics[width = .48\textwidth]{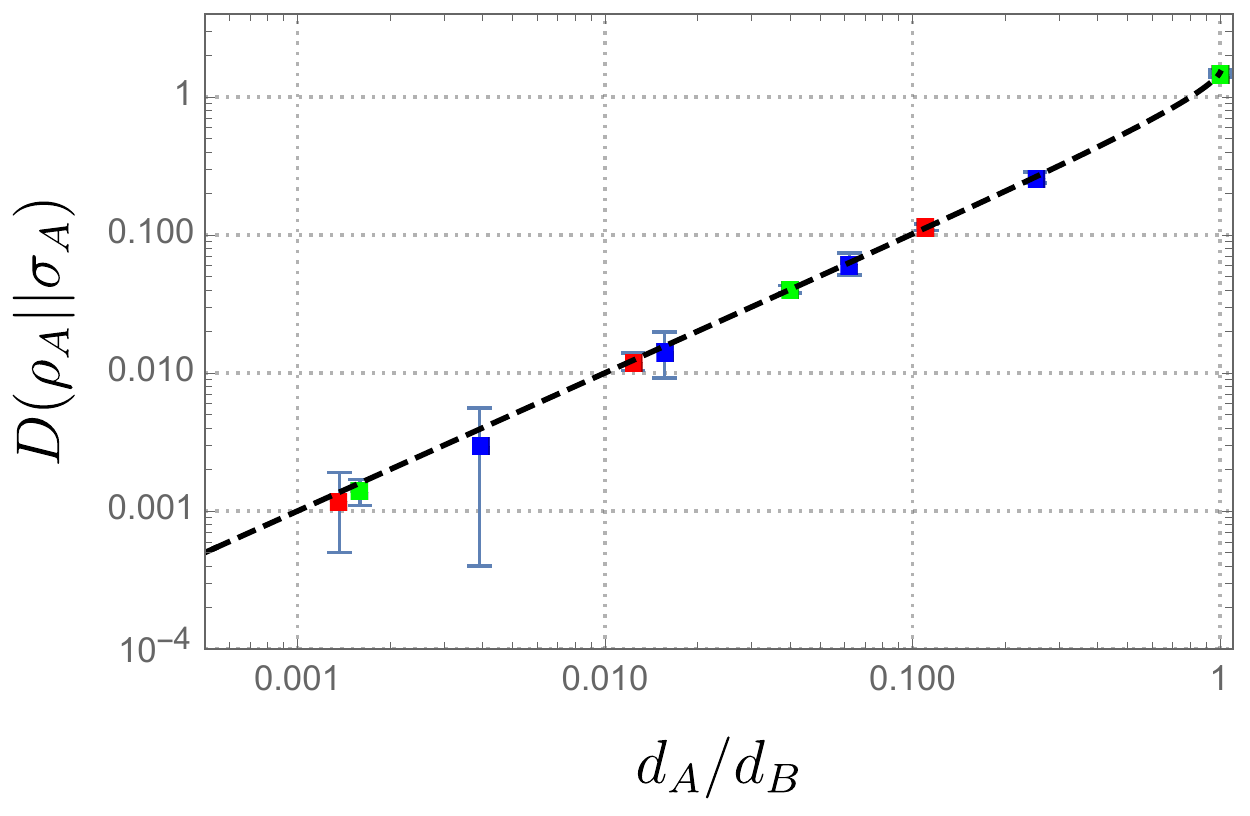}
    \caption{Comparison of equation \eqref{maineq} (dashed line) with numerics. The blue, red, and green data points are for total Hilbert space dimensions of $1024$, $6561$, and $15625$ respectively. The fluctuations in the relative entropy are clearly suppressed as the dimension is increased, signaling self averaging.}
    \label{relative_entropy}
\end{figure}

The number of non-crossing permutations stabilizing the first element is given by the Narayana number, $N_{n-1,k}$. This can be seen from the diagrams which are topological in nature
\begin{align}
    \tikz[scale=0.5,baseline=-0.5ex]{
    \draw[dashed] (0,0) -- (1,0);
    \draw (-0.2,0.)-- (-0.2,-0.35);
    \draw (1.2,-0.15)-- (1.2,0.);
    \draw[dashed,red] (2,0) -- (3,0);
    \draw[red] (2,0)-- (2,0.15);
    \draw[red] (3,0)-- (3,0.15);
    \draw[red] (1.8,-.15)-- (1.8,0.15);
    \draw[red] (3.2,-0.15)-- (3.2,0.15);
    \draw[dashed,red] (5,0) -- (6,0);
    \draw[red] (5,0)-- (5,0.15);
    \draw[red] (6,0)-- (6,0.15);
    \draw[red] (4.8,0.15)-- (4.8,-0.15);
    \draw[red] (6.2,-0.35)-- (6.2,0.15);
    \draw (1.2,-0.15)--(1.8,-0.15);
    \draw[red] (3.2,-0.15)--(3.45,-0.15);
    \draw (-0.2,-0.35)-- (6.2,-0.35);
    \draw[red] (4.5,-0.15)--(4.8,-0.15);
    \node[] at (4.,0.2) {$\cdots$};
    \draw[dashed] (1.0,0) arc (0:180:0.5);
    \draw (1.2,0) arc (0:180:0.7);
    } 
    =
        \tikz[scale=0.5,baseline=-0.5ex]{
    \draw[dashed] (0.5,0) -- (1.5,0);
    \draw[dashed,red] (2,0) -- (3,0);
    \draw[red] (2,0)-- (2,0.15);
    \draw[red] (3,0)-- (3,0.15);
    \draw[red] (1.8,-.35)-- (1.8,0.15);
    \draw[red] (3.2,-0.15)-- (3.2,0.15);
    \draw[dashed,red] (5,0) -- (6,0);
    \draw[red] (5,0)-- (5,0.15);
    \draw[red] (6,0)-- (6,0.15);
    \draw[red] (4.8,0.15)-- (4.8,-0.15);
    \draw[red] (6.2,-0.35)-- (6.2,0.15);
    \draw[red] (3.2,-0.15)--(3.45,-0.15);
    \draw (1.8,-0.35)-- (6.2,-0.35);
    \draw[red] (4.5,-0.15)--(4.8,-0.15);
    \node[] at (4.,0.2) {$\cdots$};
    \draw[dashed] (1.5,0) arc (0:180:0.5);
    } .
\end{align}
The diagrams maximizing the total number of loops are all of the non-crossing ones acting only on the $(n-1)$ red indices.
We can then reorganize the sum as
\begin{align}
    \langle{\Tr \rho_A \sigma_A^{n-1}}\rangle &= \frac{1}{(d_A d_B)^n}\sum_{k=1}^{n-1} N_{n-1,k}d_A^k d_B^{n+1-k},
    \nonumber
    \\
\end{align}
Like the R\'enyi entropies, this may also be written as a hypergeometric function
\begin{align}
    \langle{\Tr \rho_A \sigma_A^{n-1}}\rangle=d_A^{1-n} \, _2F_1\left(1-n,2-n;2;\frac{d_A}{d_B}\right).
    \label{relative_hyper}
\end{align}
Combining \eqref{renyi_hyper} and \eqref{relative_hyper}, we can unambiguously take the $n\rightarrow1$ limit to find the relative entropy\footnote{In general, the ensemble average and logarithm do not commute, requiring a further replica trick. However, these operations approximately commute for large Hilbert space dimensions. We show this explicitly in the supplemental material.}
\begin{align}
    &D(\rho_A \lvert \rvert \sigma_A) = 1+\frac{d_A}{2 d_B}+\left(\frac{d_B}{d_A}-1\right) \log
   \left(1-\frac{d_A}{d_B}\right).
    \label{maineq}
\end{align}
This is our main result. 
This formula is zero when $d_A/d_B\rightarrow 0$. This is to be expected because density matrices become indistiguishable when most of the information is ``traced away.'' The relative entropy monotonically increases with $d_A/d_B$, reaching a curious value of $3/2$ when $d_A  = d_B$. This monotonic behavior is a restatement of the monotonicity of relative entropy under the partial trace. For $d_A > d_B$, the density matrices are rank deficient leading to the formula breaking down and the relative entropy is formally infinite. 
We plot this function in Fig.~\ref{relative_entropy} and compare to numerics, finding very good agreement even for the relatively small Hilbert space dimensions that are accessible on a classical computer.

We briefly comment on the implications of \eqref{maineq} for quantum hypothesis testing, which represented a breakthrough in the operational meaning of quantum relative entropy \cite{cmp/1104248844, 2005atqs.book...28O}. Say you are given a quantum state that is either $\rho$ or $\sigma$ and you wish to determine which one you have using quantum measurements. Quantum Stein's Lemma states that the optimal asymptotic rate of error in determining which state you have is given by $e^{-D(\rho\lvert \rvert\sigma)}$ \cite{cmp/1104248844, 2005atqs.book...28O}. Thus, \eqref{maineq} tells us that if we are only given partial information about the state (access to sub-Hilbert space $A$), for a measure one set of quantum states, the error will either vanish if $A$ is larger than half the total system size or is finite and exponentially close (in the entropy) to the maximal error rate if $A$ is smaller than half the total system size.

\paragraph*{Black Hole Microstates.}

Here, we reinterpret \eqref{maineq} in the context of the AdS/CFT correspondence \cite{1999IJTP...38.1113M}. In AdS/CFT, high energy eigenstates in the boundary conformal field theory are dual to black hole microstates in the bulk because the correspondence is an isomorphism between the bulk and boundary Hilbert spaces. By black hole microstate, we therefore mean individual eigenstates of quantum gravity. Together, these microstates comprise the famous Bekenstein-Hawking entropy of the black hole \cite{Bekenstein:1972tm, Hawking:1974sw}. Precise enumerations of these microstates have been performed for special black holes \cite{1996PhLB..379...99S, 1998JHEP...02..009S}, though the general statement that the Bekenstein-Hawking entropy is truly a microscopic entropy is widely believed to be true.

Computations of relative entropy tell us how well we can distinguish different black hole microstates of similar energy i.e.~within the same microcanonical energy band\footnote{Related questions were studied in Refs.~\cite{2017PhRvD..96f6017B,2017JHEP...02..060S}.}, a notoriously difficult task that, a priori, one would expect to require knowledge of the full ultraviolet complete quantum gravity theory, such as string theory \cite{1996PhLB..379...99S}. Surprisingly, we show that this is actually possible just from semiclassical gravity, which is related to the recent surprise that the Page curve can be calculated from semiclassical gravity \cite{2020JHEP...09..002P,2019JHEP...12..063A}.

This is simplest for ``fixed-area states'' \cite{2019JHEP...05..052A,2019JHEP...10..240D}, which are holographic states where the areas of extremal surfaces in the bulk have been measured. These states have played an important role in understanding holographic entanglement entropy in the language of quantum error correction. While we first perform computations in the fixed-area state basis, we can translate these results to true energy eigenstates by noting that eigenstates are superpositions of fixed-area states with sharply peaked Gaussian distributions of width $O(\sqrt{G_N})$ \cite{2020JHEP...12..084M}. We will return to this translation in the discussion.

\begin{figure}
    \centering
    \includegraphics[width = .25 \textwidth]{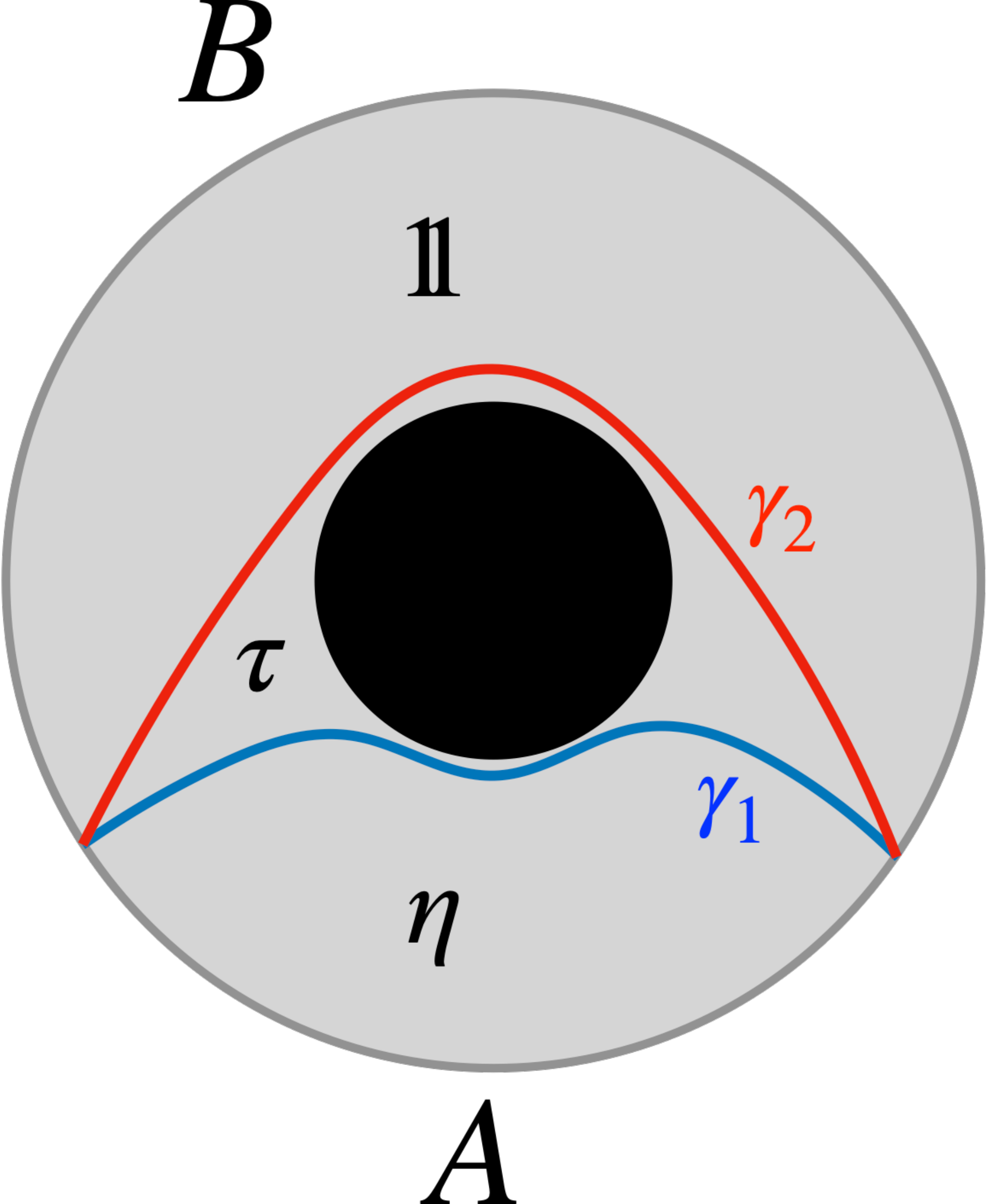}
    \caption{Depicted is a black hole geometry with the boundary partitioned into regions $A$ and $B$. There are two competing extremal surfaces, $\gamma_1$ and $\gamma_2$, that we fix the area of. When performing the replica trick, we compute the path integral on $n$ copies of this geometry. Each bulk region is labeled by the permutation element that governs how it is glued among the copies.}
    \label{RT_cartoon}
\end{figure}

While many details can be found in the original papers and illuminating follow-ups \cite{2020JHEP...03..191D,2020JHEP...12..084M,2020JHEP...11..007D,2020arXiv200803319A,2019arXiv191111977P}, we will only present what is necessary for our analysis. 
We consider states where the areas of two extremal surfaces, $\gamma_1$ and $\gamma_2$, have been measured,
as depicted in Fig.~\ref{RT_cartoon}. The two surfaces wrap the black hole horizon in topologically distinct manners\footnote{In fact, the conclusions from this section hold whenever there exist more than one extremal surface in the bulk geometry (see e.g.~Ref.~\cite{2020arXiv200803319A}). The black hole is not strictly necessary though we view it as the most physically relevant application. The existence of multiple extremal surfaces in the black hole geometry can be seen as a consequence of Ref.~\cite{2014JHEP...03..068E}. See Ref.~\cite{2013JHEP...08..092H} for explicit constructions.}. 


To compute the relative entropy between two different black hole microstates, we must compute
\begin{align}
    \Tr ( \rho_A \sigma_A^{n-1} )= \frac{\mathcal{Z}(\rho_A \sigma_A^{n-1})}{\mathcal{Z}(\rho_A)\mathcal{Z}(\sigma_A)^{n-1}},
\end{align}
where $\mathcal{Z}$ is the gravitational path integral
with the boundary conditions dictated by the argument. 
Due to nice properties of fixed-area states, the only contributions to the path integrals
come from the conical deficits that can occur at $\gamma_1$ and $\gamma_2$, leading to
\begin{align}
    \Tr ( \rho_A \sigma_A^{n-1} )= \sum_{\tau \in \mathbbm{1}\times S_{n-1}}\frac{e^{\left(D(\eta^{-1} \circ \tau) A_1 + D(\tau) A_2\right)/4G_N}}{e^{n( A_1+A_2)/4G_N }},
\end{align}
where $A_1$ and $A_2$ are the areas of the fixed surfaces. This expression is identical to \eqref{relative_sum} once we identify $d_A = e^{A_1/4G_N}$ and $d_B = e^{A_2/4G_N}$. A similar conclusion is made for $\Tr ( \rho_A^{n} )$. Therefore, the relative entropy between black hole microstates is given by \eqref{maineq}, which is UV finite because while the areas are themselves divergent, their difference is regulator independent. It is important to note that for small $A_1$, i.e.~small boundary subregion $A$, the relative entropy is nonzero, meaning the two black hole microstates are distinguishable even with very limited information about the state! The catch is that the distinguishability is non-perturbatively small in Newton's constant, $O(e^{-1/G_N})$. However, as $A_1$ approaches $A_2$, the relative entropy becomes $O(1)$. The transition from $O(e^{-1/G_N})$ to $O(1)$ occurs in an extremely tiny window when $(A_2 - A_1)/4G_N \lesssim \log 2$, roughly meaning that region $A$ contains one less qubit of information than region $B$.

In passing, we note that these results also apply to the relative entropy of two states in the Jackiw-Teitelboim gravity plus end-of-the-world brane model of black hole evaporation from Ref.~\cite{2019arXiv191111977P} in the case that the black hole is in the microcanonical ensemble.

\paragraph*{Subsystem Eigenstate Thermalization.}

Subsystem ETH is a generalization of the standard local ETH story and is significantly stronger. While ETH is a statement about local operators \cite{1991PhRvA..43.2046D,1994PhRvE..50..888S}, subsystem ETH is a statement that finite subregions appear thermal. Precisely, subsystem ETH holds when sufficiently highly excited eigenstates have reduced density matrices that are exponentially close in trace distance to some universal density matrices, such as the microcanonical ensemble \cite{2018PhRvE..97a2140D}
\begin{align}
    T(\rho_{\psi} ,\rho_{univ} ) := \left|\rho_{\psi} - \rho_{univ} \right|_1 \leq O(e^{-S(E)/2}),
\end{align}
where $e^{S(E)}$ is the density of states of the full system.
In the context of holography, the entropy scales as $O(G_N^{-1})$, so subsystem ETH means that the trace distance is nonperturbatively small in Newton's constant.

To prove this, we now invoke the quantum Pinsker inequality \cite{ohya2004quantum}
\begin{align}
    D(\rho \lvert \rvert\sigma) \geq \frac{1}{2}T(\rho, \sigma)^2.
\end{align}
We previously found $D(\rho_A \lvert \rvert\sigma_A)$ to scale as $O(e^{-1/G_N})$ for any two black hole microstates with fixed area. This implies that the trace distance is, at most, $O(e^{-1/G_N})$. The trace distance defines a metric on the space of density matrices, so if a typical state is close to a measure one set of all other states, then the universal density matrix should sit within this ball. We therefore claim that fixed-area states in all dimensions obey subsystem ETH for subsystems less than half the total system size. The violation of subsystem ETH only occurs when $(A_2 - A_1)/4G_N \lesssim \log 2$.

\paragraph*{Discussion.}

There are various interesting directions that one may take: (i) We have computed the average relative entropy between typical random mixed states. However, we have not fixed the complete distribution. It would be interesting to characterize the fluctuations in relative entropy. Higher moments of the relative entropy can be computed using the same technology that we have developed. (ii) In our applications to holography, we focused on fixed-area states. More generic states are superpositions of fixed-area states. It is important to study the relative entropy of these more generic states to verify that it is qualitatively similar. We can invoke the joint convexity of the trace distance \cite{nielsen_chuang_2010}
\begin{align}
    T\left(\sum_i p_i \rho_i, \sum_i q_i \sigma_i \right) \leq T(p_i,q_i) +\sum_i p_i T(\rho_i, \sigma_i),
\end{align}
where the $\rho_i$ and $\sigma_i$'s are fixed-area states and $T(p_i,q_i)$ is the classical trace distance between probability distributions. We have already shown that the second term on the right hand side is $O(e^{-1/G_N})$. If we assume that the probability distributions are Gaussian with equal widths but centered at fixed areas a distance at most $O(e^{-1/G_N})$ apart i.e.~within the same microcanonical window, then it is a straightforward exercise to confirm that the first term is also $O(e^{-1/G_N})$, confirming subsystem ETH\footnote{The positivity of the trace distance immediately follows from assuming the Gaussian means to be different because $T(\rho,\sigma) = 0$ only if $\rho = \sigma$.}. However, if the widths of the Gaussian distributions are different, even by an amount polynomial in $G_N$, the bound will no longer be tight.
It would be fascinating if these corrections could lead to violations of eigenstate thermalization. (iii) One of our motivations to study random states is that they should be representative of generic excited states in chaotic quantum systems. It is clearly interesting to check how accurately our results characterize real Hamiltonian systems (beyond holography). We provide numerical results for the Sachdev-Ye-Kitaev (SYK) model and for spin chains in the supplemental material. While the SYK eigenstates mimic random matrix theory, we find that chaotic spin chain eigenstates have close to, but larger, relative entropy than random states and integrable eigenstates have even larger relative entropy and much larger fluctuations. We hope to report a more systematic study in the future.

\paragraph*{Acknowledgements.}

I am grateful to Chris Akers, Hong Liu, Pratik Rath, Shinsei Ryu, Hassan Shapourian, and Shreya Vardhan for helpful discussions and comments and to Kazumi Okuyama for explaining how to simplify the functional form of \eqref{maineq}. I am supported through a Simons Investigator Award to Shinsei Ryu from the Simons Foundation (Award Number: 566166).

\paragraph*{Note Added.}
After the completion of this work, I became aware of an independent project with similar results on the computation of relative entropy \cite{Rath_unpublished}.


%

\onecolumngrid
\section*{Supplemental Material}
\beginsupplement

\section{Chaotic Eigenstates}

\begin{figure}[ht]
    \centering
    \includegraphics[width = .48\textwidth]{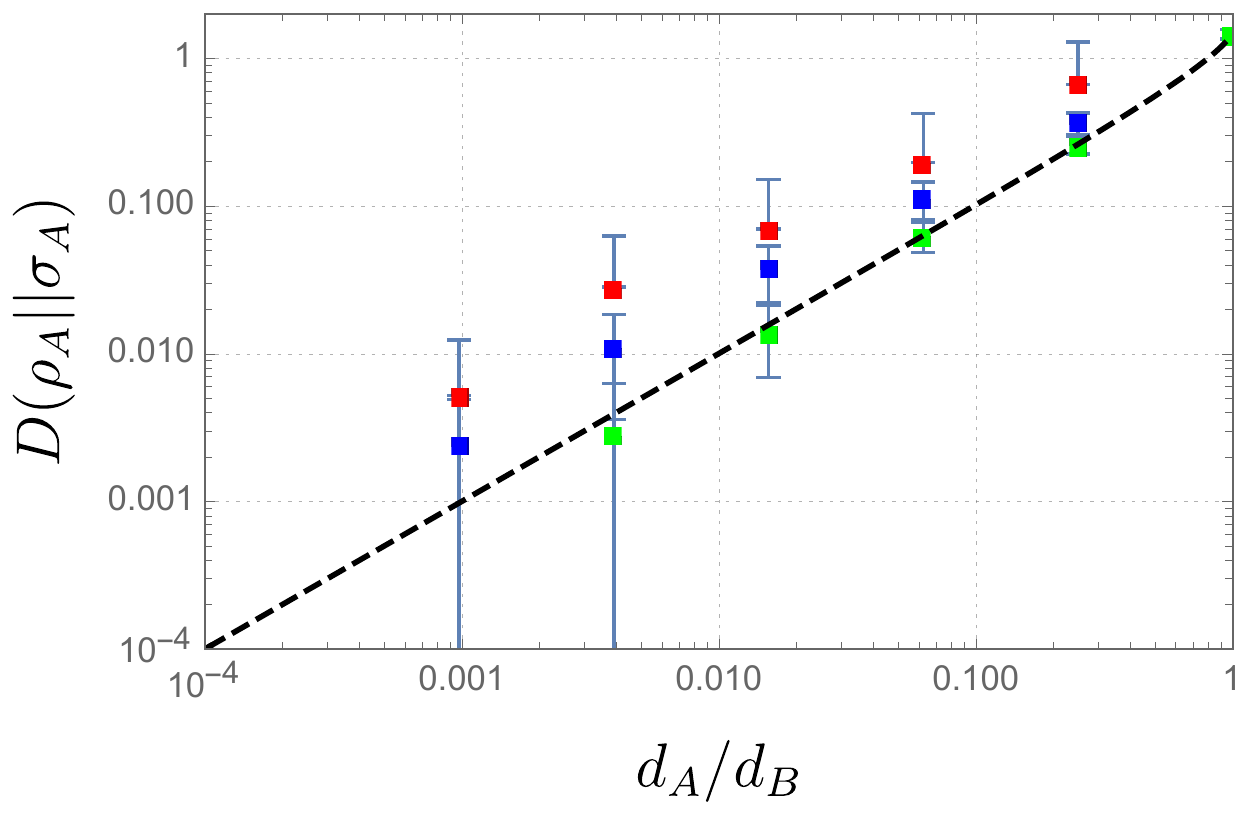}
    \caption{The relative entropy between $10^3$ random pairs of mid-spectrum eigenstates. The blue (red) data points are for the chaotic (integrable) spin chain with 12 spins and the dashed line is (23) from the main text. The green data points are for the SYK model with 20 Majorana fermions. We have omitted the lower error bars for the red data points for clarity, as they are very large and get in the way of the other data. }
    \label{chaotic_eigenstate_RE}
\end{figure}

We provide a numerical study of relative entropy between mid-spectrum eigenstates of integrable and chaotic spin chains of length $N$ with Hamiltonian
\begin{align}
    H = -\sum_{i=1}^N\left(Z_iZ_{i+1}+h_x X_i+h_z Z_i\right) ,
    \label{ham}
\end{align}
where $X$ and $Z$ are Pauli spin operators.
We take $h_x = 1$, $h_z = 0$ for the integrable limit and $h_x = -1.05$, $h_z = 0.5$ for the chaotic regime as in Ref.~\cite{2011PhRvL.106e0405B}. We also numerically study the Sachdev-Ye-Kitaev model \cite{kitaev_talk} with Hamiltonian
\begin{align}
    H = \sum_{j<k<l<m}^N J_{ijkl}\chi_j\chi_k\chi_l\chi_m,\quad \overline{J^2_{ijkl}} = \frac{6}{(N-3)(N-2)(N-1)}J^2,
\end{align}
where the $\chi_i$'s are Majorana fermions and $J_{ijkl}$ is a Gaussian random variable. The comparison between numerical data and (23) from the main text is shown in Fig.~\ref{chaotic_eigenstate_RE}. The eigenstates are chosen randomly from the middle of the spectrum. The SYK model matches very well with (23). This may be expected because the Hamiltonian is a random matrix and the SYK model is known to be closely related to low-dimensional gravitational systems. The chaotic spin chain eigenstates have relative entropy close to, but noticeably larger than, random mixed states. This is reasonable because these eigenstates are not truly random and therefore should be more easily distinguishable. It would be interesting to understand whether this is a finite size bug or a feature that holds in the thermodynamic limit. Meanwhile, the integrable eigenstates are even more distinguishable, which is consistent with their violation of the eigenstate thermalization hypothesis. Moreover, the variance in relative entropy from eigenstate to eigenstate is much larger for the integrable spin chain.

\section{Commuting of the logarithm and ensemble average}

In the main text, we computed the ensemble average of $\Tr\left[ \rho_A \sigma_A^{n-1}\right]$ and then took the logarithm to find the relative entropy. In general, the logarithm and ensemble average do not commute. However, for large Hilbert space dimensions, they approximately commute as we now show. The average of a logarithm can be computed from a further replica trick
\begin{align}
    \overline{\log \Tr\left[ \rho_A \sigma_A^{n-1}\right]} = \lim_{m\rightarrow 0}\frac{ \overline{\left(\Tr\left[ \rho_A \sigma_A^{n-1}\right]\right)^m}-1}{m}.
\end{align}
Therefore, we must compute the moments $\overline{\left(\Tr\left[ \rho_A \sigma_A^{n-1}\right]\right)^m}$. This consists of $m$ copies of the diagrams we previously considered where the ensemble averaging is allowed to connect any of the $m$ copies
\begin{align}
    {\left(\Tr\left[ \rho_A \sigma_A^{n-1}\right]\right)^m} = \tikz[scale=0.5,baseline=-0.5ex]{
    \draw[dashed] (0,0) -- (1,0);
    \draw (0,0)-- (0,.15);
    \draw (1,0)-- (1,.15);
    \draw (-0.2,0.15)-- (-0.2,-0.35);
    \draw (1.2,-0.15)-- (1.2,0.15);
    \draw[dashed,red] (2,0) -- (3,0);
    \draw[red] (2,0)-- (2,0.15);
    \draw[red] (3,0)-- (3,0.15);
    \draw[red] (1.8,-.15)-- (1.8,0.15);
    \draw[red] (3.2,-0.15)-- (3.2,0.15);
    \draw[dashed,red] (5,0) -- (6,0);
    \draw[red] (5,0)-- (5,0.15);
    \draw[red] (6,0)-- (6,0.15);
    \draw[red] (4.8,0.15)-- (4.8,-0.15);
    \draw[red] (6.2,-0.35)-- (6.2,0.15);
    \draw (1.2,-0.15)--(1.8,-0.15);
    \draw[red] (3.2,-0.15)--(3.45,-0.15);
    \draw (-0.2,-0.35)-- (6.2,-0.35);
    \draw[red] (4.5,-0.15)--(4.8,-0.15);
    \node[] at (4.,0.2) {$\cdots$};
    } 
    \hspace{.25cm}
    \tikz[scale=0.5,baseline=-0.5ex]{
    \draw[dashed] (0,0) -- (1,0);
    \draw (0,0)-- (0,.15);
    \draw (1,0)-- (1,.15);
    \draw (-0.2,0.15)-- (-0.2,-0.35);
    \draw (1.2,-0.15)-- (1.2,0.15);
    \draw[dashed,red] (2,0) -- (3,0);
    \draw[red] (2,0)-- (2,0.15);
    \draw[red] (3,0)-- (3,0.15);
    \draw[red] (1.8,-.15)-- (1.8,0.15);
    \draw[red] (3.2,-0.15)-- (3.2,0.15);
    \draw[dashed,red] (5,0) -- (6,0);
    \draw[red] (5,0)-- (5,0.15);
    \draw[red] (6,0)-- (6,0.15);
    \draw[red] (4.8,0.15)-- (4.8,-0.15);
    \draw[red] (6.2,-0.35)-- (6.2,0.15);
    \draw (1.2,-0.15)--(1.8,-0.15);
    \draw[red] (3.2,-0.15)--(3.45,-0.15);
    \draw (-0.2,-0.35)-- (6.2,-0.35);
    \draw[red] (4.5,-0.15)--(4.8,-0.15);
    \node[] at (4.,0.2) {$\cdots$};
    } 
    \hspace{.25cm}
    \dots
    \hspace{.25cm}
    \tikz[scale=0.5,baseline=-0.5ex]{
    \draw[dashed] (0,0) -- (1,0);
    \draw (0,0)-- (0,.15);
    \draw (1,0)-- (1,.15);
    \draw (-0.2,0.15)-- (-0.2,-0.35);
    \draw (1.2,-0.15)-- (1.2,0.15);
    \draw[dashed,red] (2,0) -- (3,0);
    \draw[red] (2,0)-- (2,0.15);
    \draw[red] (3,0)-- (3,0.15);
    \draw[red] (1.8,-.15)-- (1.8,0.15);
    \draw[red] (3.2,-0.15)-- (3.2,0.15);
    \draw[dashed,red] (5,0) -- (6,0);
    \draw[red] (5,0)-- (5,0.15);
    \draw[red] (6,0)-- (6,0.15);
    \draw[red] (4.8,0.15)-- (4.8,-0.15);
    \draw[red] (6.2,-0.35)-- (6.2,0.15);
    \draw (1.2,-0.15)--(1.8,-0.15);
    \draw[red] (3.2,-0.15)--(3.45,-0.15);
    \draw (-0.2,-0.35)-- (6.2,-0.35);
    \draw[red] (4.5,-0.15)--(4.8,-0.15);
    \node[] at (4.,0.2) {$\cdots$};
    } .
\end{align}
A straightforward generalization of the analysis in the main text gives
\begin{align}
     \overline{\left(\Tr\left[ \rho_A \sigma_A^{n-1}\right]\right)^m} = \frac{1}{(d_A d_B)^{mn}}\sum_{\tau \in S_m \times S_{m(n-1)}}d_A^{D((\eta^{-1})^{\times m}\circ \tau)}d_B^{D(\tau)},
\end{align}
where the $S_m$ factor of $\tau$ acts on the copies of $\rho_A$ (black lines) and the $S_{m(n-1)}$ factor acts on the copies of $\sigma_A$ (red lines). $(\eta^{-1})^{\times m}$ is the permutation element that implements the trace structure. In cycle notation,
\begin{align}
    (\eta^{-1})^{\times m} = \prod_{i = 0}^{m-1} (ni+1, ni+2,\dots, ni+n).
\end{align}
For large Hilbert space dimensions, we need to sum over the permutations that maximize $D((\eta^{-1})^{\times m}\circ \tau) + D(\tau)$.  These are non-crossing within each block of $n$ density matrices and have $D((\eta^{-1})^{\times m}\circ \tau) + D(\tau) = m(n+1)$ \cite{SIMION2000367}. Therefore, at leading order, the sum factorizes as
\begin{align}
     \overline{\left(\Tr\left[ \rho_A \sigma_A^{n-1}\right]\right)^m} \simeq \left(\frac{1}{(d_A d_B)^{n}}\sum_{\tau \in \mathbbm{1} \times S_{n-1}}d_A^{D(\eta^{-1}\circ \tau)}d_B^{D(\tau)}\right)^m =  \overline{\left(\Tr\left[ \rho_A \sigma_A^{n-1}\right]\right)}^m .
\end{align}
This immediately implies that for large Hilbert space dimensions,
\begin{align}
    \overline{\log \Tr\left[ \rho_A \sigma_A^{n-1}\right]} \simeq \log \overline{\Tr\left[ \rho_A \sigma_A^{n-1}\right]} .
\end{align}
The corrections to this formula come from the subleading permutations in $S_m \times S_{m(n-1)}$ that are not of the form $(\mathbbm{1} \times S_{n-1})^{\times m}$. An identical argument can be made for $\overline{\log\Tr\left[ \rho_A^{n}\right]}$.

Furthermore, we note that the same argument for $m = 2$ implies that the variance of ${\Tr\left[ \rho_A \sigma_A^{n-1}\right]}$ and $\Tr\left[ \rho_A^{n}\right]$ vanish in the limit of large Hilbert space dimensions. This explains the numerical results for finite dimensional systems in the main text.

\section{Hypergeometric representation}

In this section, we show (in reverse order) how the sums involving the Narayana numbers lead to standard hypergeoemtric functions.
The hypergeometric functions are defines as a power series
\begin{align}
    {}_2F_1(a,b,c;z) := \sum_{m = 0}^{\infty}\frac{(a)_m (b)_m}{(c)_m}\frac{z^m}{m!} ,
\end{align}
where $(q)_n$ is the Pochhammer symbol defined as
\begin{align}
    (q)_m = \begin{cases}
    1, & m = 0
    \\
    \prod_{i = 0}^{m-1} (q+i), & m>1
    \end{cases}.
\end{align}
The hypergeometric series terminates when either $a$ or $b$ is zero or a negative integer, in which case
\begin{align}
    {}_2F_1(a,b,c;z) := \sum_{k = 0}^{-a}(-1)^k\binom{-a}{k}\frac{ (b)_k}{(c)_k}z^k .
\end{align}
This is relevant for us because the hypergeometric functions we are interested in always satisfy this condition. In particular, plugging in the arguments for the R\'enyi entropies, we have
\begin{align}
     \, _2F_1\left(1-n,-n;2;\frac{d_A}{d_B}\right) = \sum_{k = 0}^{n-1}(-1)^k\binom{n-1}{k}\frac{ (-n)_k}{(2)_k}\left(\frac{d_A}{d_B}\right)^k
\end{align}
Note that 
\begin{align}
    (-q)_m = \begin{cases}
    1, & m = 0
    \\
    (-1)^m\prod_{i = 0}^{m-1} (q-i), & m>1
    \end{cases},
\end{align}
so
\begin{align}
     \, _2F_1\left(1-n,-n;2;\frac{d_A}{d_B}\right) = \sum_{k = 0}^{n-1}\binom{n-1}{k}\frac{ n!}{(n-k)!(k+1)!}\left(\frac{d_A}{d_B}\right)^k.
\end{align}
Redefining $k\rightarrow k-1$ and converting the factorials into binomial coefficients, we get
\begin{align}
     \, _2F_1\left(1-n,-n;2;\frac{d_A}{d_B}\right) = \sum_{k = 1}^{n}\frac{1}{k}\binom{n-1}{k-1}\binom{n}{k-1}\left(\frac{d_A}{d_B}\right)^{k-1}
\end{align}
The coefficients in this series are precisely the Narayana numbers, $N_{n,k}$. An analogous analysis can be made for $\, _2F_1\left(1-n,2-n;2;\frac{d_A}{d_B}\right)$ which was the relevant hypergeometric function for $\Tr\left[\rho_A \sigma_A^{n-1} \right]$.



Next, we perform the analytic continuation needed to compute the relative entropy. For $\Tr \left[ \rho_A^n\right]$,
\begin{align}
    \lim_{n\rightarrow 1}\frac{1}{1-n}\log\left[d_A^{1-n}\, _2F_1\left(1-n,-n;2;\frac{d_A}{d_B}\right)\right] &= \log d_A -\partial_{n} \sum_{k = 1}^n N_{n,k} \left(\frac{d_A}{d_B}\right)^{k-1}\Bigg|_{n = 1}.
\end{align}
The key point is that all Narayana numbers with $k > 2$ are proportional to $(n-1)^2$ so their derivative evaluated at $n = 1$ is trivial. Only the $k=2$ term is nontrivial. Therefore,
\begin{align}
    \lim_{n\rightarrow 1}\frac{1}{1-n}\log\left[d_A^{1-n}\, _2F_1\left(1-n,-n;2;\frac{d_A}{d_B}\right)\right] &= \log d_A -\frac{d_A}{2d_B}.
    \label{page}
\end{align}
This is Page's formula \cite{1993PhRvL..71.1291P}. For the other hypergeometric function, we have
\begin{align}
    \lim_{n\rightarrow 1}\frac{1}{1-n}\log\left[d_A^{1-n}\, _2F_1\left(1-n,2-n;2;\frac{d_A}{d_B}\right)\right] &= \log d_A -\partial_{n} \sum_{k = 1}^{\infty} N_{n-1,k} \left(\frac{d_A}{d_B}\right)^{k-1}\Bigg|_{n = 1}.
\end{align}
We take the sum to run to infinity for the purpose of analytic continuation even though the Narayana numbers are trivial for $k > n$.
We can Taylor expand the Narayana number around $n = 1$ for integer $k$
\begin{align}
    N_{n-1,k} = \begin{cases}
    0, & k = 1
    \\
    -\frac{n-1}{k(k-1)} + O(n-1)^2, & k > 1
    \end{cases}.
\end{align}
Therefore,
\begin{align}
    \lim_{n\rightarrow 1}\frac{1}{1-n}\log\left[d_A^{1-n}\, _2F_1\left(1-n,2-n;2;\frac{d_A}{d_B}\right)\right] &= \log d_A +\sum_{k = 2}^{\infty}\frac{1}{k(k-1)} \left(\frac{d_A}{d_B}\right)^{k-1}
    \nonumber
    \\
    &= \log d_A +1 +\left(\frac{d_B}{d_A}-1 \right)\log\left(1-\frac{d_A}{d_B} \right).
\end{align}
Taking the difference between this and \eqref{page} gives the relative entropy stated in the main text.

\end{document}